\documentclass[aos]{imsart}  
\usepackage{amsthm,amsmath,amssymb,natbib}

\startlocaldefs

\def\EE{\mathbb{E}}

\newtheorem{teo}{Theorem}[section]    
\newtheorem{lem}{Lemma}[section]

\endlocaldefs

\begin{document}

\begin{frontmatter}

\title{A NEW ESTIMATOR FOR THE NUMBER OF
SPECIES IN A POPULATION}
\runtitle{ESTIMATOR OF THE NUMBER OF
SPECIES }

\begin{aug}
\author{\fnms{L.} \snm{Cecconi,}
\thanksref{t1} \ead[label=e1]{cecconi@math.unifi.it}} 
\author{\fnms{A.} \snm{Gandolfi}
\thanksref{t1}
\ead[label=e2]{gandolfi@math.unifi.it}}
\and
\author{\fnms{C.C.A.} \snm{Sastri}
\ead[label=e3]{sastric@mst.edu}}
\ead[label=u1,url]{}

\runauthor{Cecconi, Gandolfi, Sastri}

\thankstext{t1}{Supported by Italian Prin project 2006010252-001.}

\affiliation{ University of Firenze and Missouri University of Science and Technology}

\address{Address of L.Cecconi and A.Gandolfi\\
Dipartimento di Matematica U. Dini,\\
 Universit\`{a} di Firenze,\\ 
 Viale Morgagni 67/A, 50134 Firenze, Italy\\ 
\printead{e1}\\
\phantom{E-mail:\ }\printead*{e2}}

\address{Address of C.C.A. Sastri\\
Department of Mathematics and Statistics\\
Missouri University of Science and Technology\\
(Formerly University of Missouri - Rolla)\\
Rolla, MO 65409\\
USA\\
\printead{e3}\\
\printead{u1}}
\end{aug}

\begin{abstract}
We consider 
the classic problem of estimating $T$, 
the total number of species in a population, from  repeated
counts in a simple random sample and look first at the
Chao-Lee estimator: we initially show that such estimator can be obtained by reconciling
two estimators of the unobserved probability, and then
develop a sequence of improvements culminating in
a Dirichlet prior Bayesian reinterpretation of the estimation problem. By 
means of this, 
we obtain simultaneous estimates of $T$, the 
normalized interspecies variance $\gamma^2$ and the
parameter $\lambda$ of the
prior. Several simulations show that our estimation method
is more flexible than several known methods we used as comparison;  the only
limitation, apparently shared by all other methods, seems to be that it cannot
deal with the rare cases in which $\gamma^2 >1$.   
\end{abstract}



\end{frontmatter}

\bigskip

\section{Introduction}

We consider the classic problem of estimating the number $T$
 of species 
 in  
a population, and, subsequentely, their distribution,
from a simple random sample drawn with replacement.
 We are interested in
the "small sample" regime in which it is likely
that not all species  have been observed.
Problems of this kind arise in a variety of settings: 
for example, when sampling fish
from a lake or insects in a forest (see, for instance, Shen, Chao and Lin (2003) \cite{SCL} on how to use
 estimates of $T$ to predict further sampling, or \cite{BMW}); 
 or when estimating
the size 
of a particular population (see \cite{BSKV});
or 
when trying to guess how many letters  an alphabet
or how many specific groups of words a language
contains (see \cite{CG})
or how many words a writer knows (see \cite{ET}); or, even, 
when determining
 how many different coins were
minted by an ancient population (Esty \cite{E}). Because
of its great interest 
this has become a classic in probability, and
there has been a great number of studies 
suggesting methods for the estimation of $T$. See, for instance,
 \cite{BF} for a review through 1993, 
\cite{GS} for some further 
details and Colwell's Estimates for software
implementing a large number of estimators. In particular, \cite{BF} calls for some
development of the Bayesian method for the 
estimation of $T$, which is the direction that we eventually
  have taken.

In this  paper we start, in fact, by analyzing one well known estimator of $T$, namely
the one by Chao and Lee (\cite{CL}). One of our results shows that
the estimator can be obtained by reconciling two estimators of the
unobserved probability $U$: one being an extended version
of Laplace's "add $\lambda$" (\cite{L}) and the other
the estimator by Turing and Good
(\cite{G}), provided that the normalized interspecies variance
$\gamma^2$ is interpreted as the inverse of the
$\lambda$. Then we proceed by developing 
simultaneous methods for estimating $T$ and $\lambda$ (or
$\gamma^2$, which is the same). 

By such methods we improve on the original Chao-Lee estimation, but
the estimators we obtain are shown by simulations to have some
serious defects. It is for this reason that we perform a more
fundamental analysis of the problem by means of a Bayesian approach.
This is based on a 
Dirichlet prior with parameter $\lambda$ on the probabilities of 
$T$ species (see \cite{Jo},
\cite{J}, \cite{G2}, 
and \cite{Z} for an historical description); the parameter
turns out to be the same as the one in Laplace's method.
The simultaneous estimation that we develop
now takes into account a posterior second moment of
the random species probabilities compared to the
classical Good Toulmin estimator for the same 
quantity (see \cite{GT}).

Let us mention that  the empirical Bayesian approach used  here 
is different from that of existing results in the literature. 
The method in \cite{MS} is, in fact, limited to 
uniform species distributions. On the other hand, the 
general Bayesian approach  in Boender and Rinnoy Kan (1987) \cite{BRK} 
starts from a prior distribution of $T$ and, 
conditionally to $T$, a uniform or Dirichlet($\lambda$) prior on
the species probability, but then introduces a (level III) prior on $\lambda$ itself
(as suggested in \cite{G3}) which in turn requires the introduction of a further
parameter (Boender and Rinnoy Kan (1987) \cite{BRK}, formulae (10) and (11)), with then
no analytical expression for the posteriors. In the end,
this direction seems to include several undetermined
choices (the prior on $T$ and the extra parameter at level III)
and no simple analytical expression of the estimators.

At the end of the paper we present some numerical tests.
Due to the inherent difficulty in finding fully published data
for this estimation we  resort to simulations and 
real tests on discovering the size of an alphabet. 
The tests seem to indicate that 
 the new estimator of $T$ is
more flexibile than existing ones and thus preferable, in the sense
that the performance of all estimators seem to greatly depend on 
the normalized  variance $\gamma^2$, and the new estimator is the only one
able to perform rather
well for all values of $\gamma^2 \in [0,1]$.
In our method, the only constraint is that   $\lambda \geq 1$,
which is $\gamma^2 \leq 1$, which is imposed in order to ensure
 convergence of the prior;
this, in turn, imposes a mild limitation on
the  populations to which the method can be applied, since
$\gamma^2$ can, for some peculiar population,
exceed $1$; on the other hand, such populations are likely to be 
quite unusual and, in addition,
all other existing estimators seem also to fail on samples taken
from them.

In section 2 we review in detail some known
estimation methods of interest in deriving our results; in 
section 3 we derive some relations between known estimators and
our first improvements; in section 4 we develop the Bayesian
method and define our final estimator; in section 5
we
give  estimates of the species probabilities from, for both the observed and
the unobserved ones; from these, we indicate how to generate
confidence intervals for $T$ by means of resampling;  finally, in section 6 we
present some simulations revealing a rather good performance of
our new estimator and also very adequate results of
 the confidence intervals. All  detailed mathematical
proof are deferred to the Appendix.

\medskip
\section{Some known estimators of $T$ and related quantities}

We start with some notation. Assume that 
the  population from which the sample is drawn has a total of $T$ species
(which we sometimes will call  
states) having proportions $p_1, p_2, \cdots , p_T.$; 
 and that in a sample
$x_1, x_2, \cdots , x_n$ of size $n$ there are $N$ observed species.
For $i = 1, \cdots , 
T$, let $m_i$ be the number of observations of the species 
$i$ in the sample, 
so that 
$\sum_{i=1}^N m_i= n.$ We assume that  the $m_i$'s are given one of the 
possible orders in which 
$m_1 \geq m_2 \dots, m_N \geq 1$
 and $m_i = 0$ for $i = N+1, \dots, T$.
Also, for $j = 1, \cdots , n$, let $n_j$ be the 
prevalence of $j$, which is to say the number of 
species 
observed exactly $j$ times, so that $\sum_{j=1}^n n_j= N.$ 
Next, 
let $L_n(i)= m_i/n$ be the empirical frequency of species 
$i$, so 
that $C = \sum_{i:L_n(i)>0}p_i$ is the coverage, i.e, the 
total 
probability of  the observed species, and 
$ U =1-C= \sum_{i:L_n(i)=0}p_i$ is the unobserved 
probability. We are interested in the estimation of $T$ from
the prevalences.

The estimation of $U$ has also been studied intensively
(see, for instance, \cite{M} and \cite{LMP}).
 In fact, it is possible to turn the estimation of $U$ into a  simplified
version of our original problem by assuming
 that there are $N+1$ 
species, the $N$ observed ones and the "new" species with probability
$U$; the main issue becomes then  the estimation of  the probabilities
of the various species and especially for the new one.
For this and other reasons that we shall see, 
the estimations of $T$ and $U$ are closely intertwined
(even the title of \cite{E85} points to this relation).

The first attempt to estimate $U$ can be 
extracted from Laplace (see \cite{L} and \cite{OSZ})
who suggested an "add-one" estimator: this consists in adding one
to the number of observations of each species plus an
additional one for  the "unobserved" species. 
In an extended version, which can be named "add $\lambda$", one can add some positive value $\lambda$
 to each species' number of observations (including the
unobserved one): 
an estimate of the
probability of each observed species $i$
is then $\widehat{p_i}=\frac {m_i+\lambda}{\lambda+\sum_{i \geq 0} (m_i+\lambda)}
=\frac {m_i+\lambda}{n+(N+1)\lambda}$
and the
estimate of the unobserved probability becomes 
$\widehat{U}_{L, \lambda}= \frac {\lambda}{n+(N+1)\lambda}.$ 

\medskip

With a seemingly completely different method, Turing and
Good (see \cite{G}) proposed another estimator
of $U$. Recall that $n_1$ is the number of species observed exactly once and $n$ 
the size of the sample; then the Turing-Good estimator for $U$ is
 some minor modification of:
$$ \widehat{U}_{TG} = \frac {n_1}{n}. $$
A plausible rationale for this estimator is that while for species 
observed at least twice the empirical frequency is already becoming 
stable and very likely close to the corresponding probability, 
 species observed only once
are likely to be randomly selected representatives of the collection of the yet 
unobserved species. A more sound mathematical derivation is in Good (\cite{G}),
in which also a "`smoothing"' of the $n_i$'s is proposed.

\smallskip

Other methods to estimate $U$ have been developed, and in particular we refer to
\cite{LMP} for a Bayesian method based on the general class of Gibbs-type priors (see also \cite{P} and the other references in \cite{LMP} for the definition and properties of such priors). This class contains several known families of priors as particular cases and each such family is based on one or more parameters, which need to be further estimated. In \cite{LMP}, for instance, a maximum likelihood estimator is used.
Another recent advance 
 appears in Orlitsky et al (\cite{OSZ}),
in which a quantity is introduced, called attenuation,
that measures the effectiveness of the estimation of
$U$
 as the sample gets larger; the performance of an estimator
is compared to the maximum probability of the observed 
prevalences and asymptotically
very good estimators are determined.

We are going to base our work here on a preliminary estimation of $U$.
 It is conceivable that within the wide class of proposed estimators
 of $U$ some would improve the results that we get; however, 
  we focus on the unsmoothed
Turing-Good estimator since it is more direct and simple, while
still allowing us to achieve very satisfactory results.

\smallskip

Getting back to the estimation of $T$, there are several parametric
 methods based on assuming some structure
 of the species distribution;  
 for instance, an
 estimator devised for the uniform
 case, in which the probabilities of all species are
 assumed to be  the same is 
 the Horvitz-Thompson
 $$ \widehat{T}_{HT} = \frac{N}{1-U},$$
 (see \cite{LR} and Bishop, 
Fienberg and Holland  (1975) \cite{BFH})
 and then $U$ can be further estimated, for instance
  by the unsmoothed Turing-Good method, to get
\begin{equation}\label{THTTG}
 \widehat{T}_{HTTG} = \frac{N}{1-\widehat{U}_{TG}}
=\frac{nN}{n-n_1}
\end{equation}
see \cite{DR} and \cite{BS}.  Esty \cite{E85} improves this estimate by assuming 
a negative binomial prior with parameter $k$ to get
\begin{equation}\label{E}
 \widehat{T}_{HTTG} = \frac{N}{1-\widehat{U}_{TG}}+ \frac{n
\widehat{U}_{TG}}{(1-\widehat{U}_{TG})} \frac{1}{k},
\end{equation}
then providing some ad hoc guess for $k$ (in some cases, $k=2$).

As to nonparametric methods, 
Harris \cite{H},
Chao \cite{C} and Chao \& Lee \cite{CL} have proposed some such 
estimators, of which the most reliable ones seem to be 
those proposed in \cite{CL}. In our notation these amount to
\begin{eqnarray}\label{eq1}
 \widehat{T}_{CL}(\widehat{\gamma}) &=&\frac{N}{1-\widehat{U}_{TG}} + \frac{n
\widehat{U}_{TG}}{(1-\widehat{U}_{TG})} \widehat{\gamma}^2, 
\end{eqnarray}
with $\widehat{\gamma}^2$ an estimate - for which Chao \& Lee make 
two proposals - 
of the normalized variation coefficient  of the $p_i$'s. In fact,
assume that $p$ is a random variable uniformly distributed on 
the $T$ population probabilities $p_1, \dots, p_T$;, then its average
is
   $$\bar{p}=\frac{1}{T} \sum_{k=1}^{T} p_k =\frac{1}{T},$$
and its normalized variation coefficient is 
\begin{equation}\label{gamma}
	\gamma^2=\frac{Var(p)}{[\mathbb{E}(p)]^2}= T \sum_{k=1}^{T} (p_k-\bar{p})^2
	=T \sum_{k=1}^{T} p_k^2 - 1.
\end{equation}
Next, Chao and Lee proceed by using an estimate of 
Good and Toulmin 
\begin{equation}\label{Toulmin}
\sum_{k=1}^{T} p^2_k \approx \widehat V_{GT}=\sum_{j \geq 1} \frac{j(j-1)n_j}{n(n-1)}
\end{equation}
and using one preliminary estimate for $T$,  \eqref{THTTG} for 
instance, to obtain
\begin{displaymath}
\widehat{\gamma}^2= \max \Big( \frac{nN}{n-n_1} \sum \frac{j(j-1)n_j}{n(n-1)}-1,0 \Big) \label{gammaCL}.
\end{displaymath}
Note that the work by Chao and Lee can be considered as a further improvement 
over the results by Esty.
However, Chao and Lee make a rather direct use of a preliminary guess for $T$
and we think their method is too sensitive to  errors in such preliminary evaluation.
In the next section we start discussing some possible improvements.

\medskip

\section{Preliminary results on new estimators}

(I) We first consider \eqref{eq1} and \eqref{gamma}
as equations in the unknowns $T$ and $\gamma^2$ and search for 
 simultaneous solutions $T \geq N$ and $\gamma^2 \geq 0$.
Since in some simple examples the unique solution gives $\gamma^2 < 0$, we
 consider the solutions $T_1(\widehat \gamma_1^2)$ and $\widehat \gamma_1$ of the problem
\begin{eqnarray} \label{simpProb}
 T &=&T(\gamma^2)= \frac{N}{1-\widehat{U}_{TG}} + \frac{n
\widehat{U}_{TG}}{(1-\widehat{U}_{TG})} {\gamma}^2  \\
\widehat \gamma^2&=& \arg \inf_{ \gamma^2 \geq 0} \Big| \gamma^2
 -(T  \widehat V_{GT} -1)\Big|,
\end{eqnarray}
with $\widehat V_{GT}$ as in 
\eqref{Toulmin}.
On letting $u= \widehat{U}_{TG}$ and $v=\widehat V_{GT}$ for brevity, the function to
minimize becomes
$$
(1-u+nuv)  \gamma^2 + 1-u-Nv;
$$
note that $(1-u+nuv) \geq 0$ since $u \leq 1$, so that the 
solutions of \eqref{simpProb} are
\begin{displaymath}
\widehat\gamma^2_1 = \left\{ \begin{array}{ll}
0 & \textrm{if $0< u \leq 1-N v$} \\
\frac{Nv-1+u}{1-u+n u v} 
= \frac{N\widehat V_{GT}-1+\widehat{U}_{TG}}{1-\widehat{U}_{TG}+n \widehat{U}_{TG} \widehat V_{GT}} & \textrm{if $ 1-N v< u $}
\end{array} \right.
\end{displaymath}
and $\widehat T_1=T_1(\widehat\gamma_1^2)$. 

Some tests described in section 6 show that $\widehat T_1$ performs better
 for non uniform populations than the original Chao-Lee estimate, but has 
 too large a variance.

\smallskip

(II) 
Next we compare two 
 estimators  of $U$, the unsmoothed Turing-Good and the following modified version of the  "add $\lambda$": assume the number $T$ of species is known and
 add $\lambda$ to each of the  frequencies
of all the $T$ species, not just to that of those arbitrarily labelled through $N+1$. This
would give
\begin{eqnarray}
\widehat{p}_k(\lambda) & = & \frac{m_k+\lambda}{T\lambda+n} \quad \textrm{ per } k=1 \dots N \nonumber \\
\widehat{p}_k(\lambda) & = & \frac{\lambda}{T\lambda+n} \quad \textrm{ per } k=N+1 \dots T 
\nonumber \\
\widehat{U}_{\lambda}  & = & \frac {(T-N) \lambda}{T  \lambda+n} \nonumber
\end{eqnarray}
since there are $T-N$ unobserved species.
Now, we can hope to reconcile the extended
"add $\lambda$" and the unsmoothed Turing-Good estimators by requiring 
that they assign the same value to $\widehat U$. This
amounts to solving
\begin{equation}\label{reconcil}
\frac {(T-N) \lambda}{T  \lambda+n}=\widehat{U}_{TG} = \frac {n_1}{n}.
\end{equation}
Solving for $T$ we get
\begin{equation} \label{T_lambda}
\widehat{T}_{\lambda}= \frac{N+n \widehat{U}_{TG}/ \lambda}{1-\widehat{U}_{TG}}=
n \frac{N+n_1/ \lambda}{n-n_1}.
\end{equation}
Quite surprisingly, we have obtained
\begin{lem}
The only value of $T$ for which the extended "`add $\lambda$"' and 
the Turing-Good estimators of $U$ coincide, is 
the Chao-Lee estimator $T_{CL}(\gamma)$ with
$\gamma^2=1/ \lambda$. From now on we will assume this equality
and mostly refer to the parameter $\lambda$.
\end{lem}

\smallskip

(III) The relation found in (II) suggests that \eqref{simpProb} can
be seen as a first moment estimate: 
\begin{equation}\label{firstmom}
\sum_{k=N+1}^T \widehat{p}_k(\lambda)= \widehat U_{TG},
\end{equation}
so that one can hope to derive $\gamma^2$ from a second moment relation. The form
is suggested by (I), considering the meaning of $\widehat V_{GT}$:
\begin{equation}\label{secondmom}
\widehat \lambda_2= \arg \inf_{\lambda \geq 0} |\sum_{k=1}^{T} \widehat{p}_k(\lambda)^2
 - \widehat V_{GT}|.
\end{equation}
The solutions $\widehat T_2(\widehat \lambda_2)$ and $\widehat \lambda_2$
of \eqref{firstmom} and \eqref{secondmom}, together with
$\widehat \gamma^2= \widehat \lambda_2^{-1}$, give new estimators; although this
seems to improve the estimation in some cases, it does appear to have significant
flaws, as shown in the simulations reported in tables 1-3.

\bigskip

\section{ The Bayesian interpretation}

To further improve the above estimate, we need 
 to understand more about the
 "add $\lambda$" estimator.
It turns out, as was probably known already to Laplace, that the
probability estimation according to the 
"add $\lambda$" method is nothing  but the average species
probability under the Bayesian posterior on
probability distributions on $T$ species
$$ \Sigma_T = \{p = (p_1, p_2, \cdots, p_{T}),
p_i\geq 0, 
\sum_{i=1}^{T}p_i = 1 \},$$ given the sample, 
with a  single parameter Dirichlet prior $\rho_{0,T,\lambda}$,
i.e. a prior  with density 
$c \prod_{i=1}^T p_i^{ \lambda -1}$ for some constant $c$ and $\lambda \geq 1$. 
With likelihood
$$ \mu(x) = c \prod_{j=1}^n p^{\lambda -1}_{x_j} = 
c \prod_{i=1}^T p_i^{m_i+\lambda -1} $$
 the posterior becomes
\begin{eqnarray}
\rho_{n,T,\lambda}(d\mu) &=& \frac{\mu(x)\rho_{0,T,\lambda}(d\mu)}
{\int_{\Sigma_T}\mu(x) \rho_{0,T,\lambda}(d\mu)}\\
&=& 
\rho_{n,T,\lambda}(d \mu)= \frac{1}{Z_{\Lambda}} {\bf 1}_{\Sigma_T}
\prod_{i=1}^T p_i^{m_i+ \lambda -1}dp_1 \dots dp_T.\nonumber
\end{eqnarray}
where  
$Z = 
\int_{\Sigma_T} 
p_1^{m_1+ \lambda -1} \cdots p_N^{m_N+ \lambda -1}p_{N+1}^{ \lambda -1}
\cdots p_{T}^{ \lambda -1} dp_1\cdots dp_{T}$ 
(note that the constant terms have been cancelled).

By standard integration using the gamma function
(see Appendix 1), we find that the  average species probability under the posterior is:

\begin{displaymath} \label{e:expectedy1}
E_{\rho_{n,T, \lambda}}(y_i) = \left\{ \begin{array}{ll}
\frac{m_i +\lambda}{T \lambda+n} & \textrm{if $i=1, \dots, N$} \\
\frac{\lambda}{T \lambda+n}    & \textrm{if $i=N+1, \dots, T$}
\end{array} \right.
\end{displaymath}
as claimed. 
This remark,  together with our reconcilation Lemma in (I) above, 
indicates that we are taking a new step in the development which
brought us from \eqref{THTTG} to \eqref{E} and then to \eqref{eq1}
by assigning now two other meanings for $\lambda^{-1}
= \gamma^2$, namely  that of the add constant in a generalized Laplace
method and that of the constant in a Dirichlet prior.

\medskip

The Bayesian interpretation of $\widehat p_k$ also suggests a modification of the
second moment minimization \eqref{secondmom}. Recalling that now 
$\lambda \geq 1$ we have: 
$$
\widehat \lambda= \arg \inf_{\lambda \geq 1} |f(\lambda)|
$$
 with
\begin{eqnarray} \label{e:f}
 f(\lambda)&=&   \widehat V  - \sum_{k=1}^{T} (\EE_{\rho_{n,T,\lambda}}(p_k^2))
  \nonumber \\
&=&  \frac{\sum_{j>0} j^2n_j-n}{n(n-1)}
- \frac{2 n \lambda+\lambda+n \lambda(\lambda +1)\frac{N\lambda+n_1}{n-n_1}}{[n\frac{N\lambda+n}{n-n_1}+1][n\frac{N\lambda+n}{n-n_1}]} \nonumber
\end{eqnarray} 
where  $\widehat  T_{\lambda}$ has been taken as in \eqref{T_lambda}
and the calculation is carried out in Appendix 1.
In Appendix 2 we show the  function $f(\lambda)$ has two singularities
$\beta_2 < \beta_1 =-\frac{n}{N}<0$ and two zero's, the interesting one being
\begin{eqnarray} \label{e:lambda2}
\lambda_2 & = & \frac{1-u-v+uv-uvn}{Nv+u-1}.
\end{eqnarray}

The minimization depends on the sign of 
$f(\lambda)$ for large $\lambda$ which in turn depends on
the sign of 
$(\lambda_2-\beta_1)$. Since $f(\lambda)$ is
increasing for $\lambda \geq 1$, if the limit for large
$\lambda$ is
negative, then the only reasonable value we can
assign is $\infty$, else there is a real solution for 
the minimization problem above: note that 
if $\lambda_2 \leq 1$ then we are forced to take 
$\widehat{\lambda}=1$.
It is thus shown in Appendix 2 that the minimization above yields the estimator
\begin{displaymath}
\widehat{\lambda} = \left\{ \begin{array}{ll}
 1 & \textrm{if $\beta_1 < \lambda_2$ and $1 \geq \lambda_2$, i.e. $\frac{2-v(N+1)}{2-v+vn} \leq u \leq 1-v$} \\
\lambda_2  & \textrm{if $\beta_1 < \lambda_2$ and $ \lambda_2 \geq 1$, i.e. $1-N v < u \leq \frac{2-v(N+1)}{2-v+vn}$} \\
\infty    & \textrm{if $\lambda_2 \leq \beta_1$, i.e. $0 \leq u \leq 1-N v$}.
\end{array} \right.
\end{displaymath}
From \eqref{T_lambda} we get the following estimator of $T$:
\begin{displaymath}
\widehat{T}_{\widehat \lambda}= \frac{N+n \widehat{U}_{TG}/ \widehat\lambda}{1-\widehat{U}_{TG}} = \left\{ \begin{array}{ll}
n \frac{N+n_1/ (\lambda_2 \vee 1)}{n-n_1} & \textrm{if $\beta_1 < \lambda_2$} \\
\frac{n N}{n-n_1}    & \textrm{if $\lambda_2 < \beta_1$}.
\end{array} \right.
\end{displaymath}
or, alternatively,
\begin{displaymath}
\widehat{T}_{\widehat \lambda} =\left\{ \begin{array}{ll}
 \frac{N+n u}{1-u} & \textrm{if $\frac{2-v(N+1)}{2-v+vn} \leq u \leq 1-v$} \\
 \frac{N-N v -n u}{1-u-v+uv-uvn} & \textrm{if $1-N v \leq u \leq \frac{2-v(N+1)}{2-v+vn}$} \\
\frac{N}{1-u}    & \textrm{if $0 \leq u \leq 1-N v$}.
\end{array} \right.
\end{displaymath}

Clearly, $\widehat{T}_{\widehat \gamma^2}$ is not necessarely an integer while $T$ is such, and 
we round it to the nearest integer. Notice that when $\widehat \lambda= \infty$
we get $\widehat T_{\widehat \lambda}= \widehat T_{HTTG}$.

\medskip
\section{Estimate of species distribution and confidence intervals for $T$}

Since we now have an estimate for both the parameters
$T$ and $\lambda$, we can use  the posterior average probability of each species 
as an estimate of 
 the species probabilities. For
the observed species, i.e. for  $i=1, \dots, N$, this amounts to
\begin{equation} \label{EstObsSpec}
\widehat p_i= E_{\rho_{n, \widehat {T}_{\widehat\lambda},\widehat\lambda}}(y_i) =
\frac{m_i +\widehat\lambda }{ \widehat {T}_{\widehat\lambda} \widehat\lambda+n}
=\frac{(m_i +\widehat\lambda)(1-\widehat U) }{ n+N\widehat\lambda  } .
\end{equation}
This expression is correct also for $\widehat\lambda = \infty$
in which case all species are estimated to have probability
$(\widehat T)^{-1}$. Also note that these values are close to the unbiased estimator
$m_i/n$ of the probability of the $i$-th species and can be seen as
 a mixture of the Laplace add-$\lambda$ and Turing-Good estimators
 since they are obtained by adding $\lambda$ to the frequency $m_i$ of  the
$N$ observed species (recall that $n= \sum_{i=1}^N m_i$), but only after having 
assigned the probability $\widehat U$ to the event that we will observe a new
species;  the estimate of each of the $N$ species is then
reduced by the factor $1-\widehat U$ to compensate for this and, in fact,
$( \widehat{T}_{\widehat\lambda}-N)  \frac{\widehat\lambda(1-\widehat U) }{ n+N\widehat\lambda+ \widehat U}= \widehat U$. This is likely to be a 
sensible way to make
the attenuation of the Laplace estimator (see \cite{OSZ}) finite.
An alternative description of our estimator is then completed by using
 the previously estimated value of $\lambda$.

A simple approach for the unobserved species would be 
to uniformly split the probability $\widehat U$ among the 
$ \widehat {T}_{\widehat\lambda} - N$ unobserved species and by
the reconcilation method in 
 \eqref{reconcil} and \eqref{T_lambda} this would give
 $\frac{\widehat U}{\widehat {T}_{\widehat\lambda} - N}=
 \frac{\widehat\lambda }{ \widehat {T}_{\widehat\lambda} \widehat\lambda+n}
=\frac{\widehat\lambda(1-\widehat U) }{ n+N\widehat\lambda} $.
On the other hand, notice that,
since one can read \eqref{firstmom} as
$1-\sum_{k=1}^N \widehat{p}_k(\lambda)=1-  \widehat U_{TG}$,
 the reconciliation method never used the moments of the $p_i$'s for
$i >N$; therefore, 
 we have some freedom in assigning
the estimated values of the $p_i'$s for $i>N$. These values
can then be estimated by taking
 into account the meaning of
$\lambda^{-1}=\gamma^2$  as normalized species variance, or of some related quantities;
we could then 
assign probabilities to the
unobserved species to achieve the estimated normalized variance
$\widehat \gamma^2$ or to achieve some related equality. For simplicity we will actually focus on
$\sum_{k=1}^N p_k^2$ and its estimator $\widehat V$.
This is a valid approach except when
$u < 1-Nv$, in which case $f(\lambda) <0$ and 
$\widehat V$ turns out to be too small to be a reasonable estimate
of $\sum_{k=1}^N p_k^2$; in that case we replace  $\widehat V$ with 
$\sum_{k=1}^{\widehat T_{\widehat \lambda}} \EE_{\rho_{n,\widehat T_{\widehat \lambda},\widehat \lambda}}(p_k^2)$.
Clearly
$$
 \sum_{k=1}^{N} (\EE_{\rho_{n,T,\lambda}}(p_k))^2
\leq\widehat V \vee \sum_{k=1}^{N} \EE_{\rho_{n,T,\lambda}}(p_k^2)
$$
by Jensen's inequality, and thus we require that the estimates $\widehat p_k$ of the
probabilities of the unobserved species
satisfy:
$$\sum_{k=N+1}^{\widehat T_{\widehat \lambda}} (\widehat p_k)^2
=\left( \widehat V \vee \sum_{k=1}^{\widehat T_{\widehat \lambda}} \EE_{\rho_{n,\widehat T_{\widehat \lambda},\widehat \lambda}}(p_k^2) \right) - \sum_{k=1}^{N} 
(\EE_{\rho_{n,\widehat T_{\widehat \lambda},\widehat \lambda}}(p_k))^2=: \tilde V
$$

We can  use any two parameter distribution, such as for instance 
$p_i=c \alpha^{i-N}$ for $i=N+1, \dots, \widehat {T}_{\widehat\lambda}$, and
insist that 
\begin{equation} \label{FirstEquatParUnobs}
\sum_{i=N+1}^{\widehat {T}_{\widehat\lambda}} p_i= \widehat U_{TG}
\end{equation}
and 
\begin{equation} \label{SecondEquatParUnobs}
\sum_{i=N+1}^{\widehat {T}_{\widehat\lambda}} p_i^2= \tilde V.
\end{equation} 
Solving for $c$ and $\alpha$  gives the estimated unobserved
probabilities $\widehat p_i= p_i(c, \alpha)$, which are used in the
simulations of section 6 below to generate confidence intervals by resampling.

It is easily seen that if $T > > N$ then 
$$
\alpha(1-\alpha) \approx u/v
$$
and 
$$
c \approx \frac{u(1-\alpha)}{\alpha}.
$$

\medskip

\section{ Simulations}

In this section we present numerical simulations and tests of the performance of several estimators 
compared to those we have developed here. Tables 1-4 present the analysis of several
populations increasing values of $\gamma^2$. Tables 5-6 
present some real tests based on discovering the number of letters in an alphabet
from a long text. In table 7 we compute confidence intervals using a resampling based on
the reconstructed species' probabilities as described in section 5 above.

The estimators compared in tables 1-6 are
$\widehat T_1$, $\widehat T_2$ and $\widehat T_{\widehat \lambda}$
defined here, then $\widehat T_{THTTG}$ from \eqref{THTTG},
 $\widehat T_{CL}$ from \eqref{eq1}, the Jackknife estimator with 
 optimal parameter $\widehat T_{JK}$ from \cite{BO}
 (// indicates numerical errors due to small denominators),
 and $\widehat T_{+1}$ which is our (or the Chao-Lee) estimator with
 $\gamma^2=1$.

In tables 1-4 each
population is generated from $T$ i.i.d. random variables, normalized to sum to $1$; 
the resulting $\gamma^2$ is determined as normalized interspecies variance; 
$1000$ simple random samples of size $n$ are then generated; finally, mean,
SD and mean square error are computed for each
estimator. 

Tables 5 and 6 test the letter content of 
some passages in English and Italian in order to detect the number
of letters in each alphabet. Each table shows 
the results of taking $1000$ samples of about
$9000$ letters each from the indicated texts. 

\smallskip

The conclusion that can be drawn from these tests is that 
estimator performances are seen to depend on $\gamma^2$, with the
$\widehat {T}_{\widehat \lambda}$ presenting a consistent low value of
the MSE as long as $\gamma^2 \in [0,1]$. Therefore, 
 $\widehat {T}_{\widehat \lambda}$ has the flexibility to 
adapt to the different values of the interspecies variance.
In table 1, in fact, $\gamma^2 \approx 0$
and the best estimators turn out to be $\widehat T_{THTTG}$
and $\widehat T_{CL}$ (in which clearly $\gamma^2$ gets appropriately 
estimated), but all the estimators defined in the present paper perfom equally 
well. In the less uniform population in table 2, Jackknife and 
$\widehat T_{\widehat \lambda}$ show the best performances; 
and in table 3 where $\gamma^2 \approx 1$, the best estimator
turns out to be $\widehat T_{+1}$, while $\widehat {T}_{\widehat \lambda}$ has only 
a slightly worse performance. Note that 
$\widehat T_1$ and $\widehat T_2$ show a very poor performance
in table 2 and 3.

Finally, table 4 shows an extremely skewed population, 
with $\gamma^2$ very large, for which no estimator works properly. The
reason for $\widehat T_{\widehat \lambda}$ is that  convergence of the prior 
imposes
$\gamma^{-2}=\lambda \geq 1$.

Even in the alphabet test the performance of $\widehat T_{\widehat \lambda}$
turns out to be overall best.

Table 7 shows some simulations about confidence intervals for $T$ based on 
samples of size $n=400$ computed from $\widehat T_{\widehat \lambda}$
by estimating the species probabilities $p_k$ as described in 
section 5 and then resampling $1000$ times from the estimated
population distribution. This process is repeated $100$ times
and table 7 indicates, for the populations of tables 1-3 respectively,
the percentage of times the confidence intervals hits the true 
value of $T=1000$ and the average size of the confidence interval.

The hitting percentage comes out remarkably well, due to the good approximation
of the true population distribution by the estimated one.

\begin{table}[ht]
\centering

\begin{tabular}{|c || c|c|c || c|c|c || c|c|c ||}
\hline   
$T=1000$ & \multicolumn{3}{|c||}{$n=500$} & \multicolumn{3}{|c||}{$n=1000$} & \multicolumn{3}{|c||}{$n=2000$} \\
\cline{2-10}
                           & mean & std & MSE & mean & std & MSE & mean & std & MSE \\
\hline
\hline
$\widehat{T}_{TG}$         & 994  & 79  & 79  & 999  & 36  & 36  & 997  & 16  & 16 \\
$\widehat{T}_{CL}$         & 1010 & 86  & 87  & 1009 & 42  & 43  & 1000 & 18  & 18 \\
$\widehat{T}_{JK}$         & 1068 & 96  & 117 & 1223 & 84  & 239 & 1117 & 165 & 203\\
$\widehat{T}_{+1}$         & 1759 & 157 & 775 & 1580 & 73  & 585 & 1309 & 32  & 311\\
$\widehat{T}_1$            & 1003 & 82  & 82  & 1005 & 39  & 40  & 1000 & 18  & 18 \\
$\widehat{T}_2$            & 1017 & 83  & 86  & 1010 & 38  & 40  & 1027 & 54  & 60 \\
\hline
$\widehat{T}_{\lambda}$    & 1087 & 193 & 212 & 1026 & 60  & 66  & 1002 & 20  & 20 \\

\hline
\end{tabular}
\bigskip

\caption{Uniform population: $p_i$'s $\sim N(0,1)$, $\gamma^2 \approx 0.009$.}
\end{table}
\bigskip
\bigskip

\begin{table}[ht]
\centering

\begin{tabular}{|c || c|c|c || c|c|c || c|c|c ||}
\hline   
$T=1000$ & \multicolumn{3}{|c||}{$n=500$} & \multicolumn{3}{|c||}{$n=1000$} & \multicolumn{3}{|c||}{$n=2000$} \\
\cline{2-10}
                           & mean & std & MSE & mean & std & MSE & mean & std & MSE \\
\hline
\hline
$\widehat{T}_{TG}$         & 781  & 59  & 226 & 816  & 30  & 186 & 858  & 15  & 142 \\
$\widehat{T}_{CL}$         & 808  & 72  & 205 & 847  & 40  & 158 & 893  & 20  & 109 \\
$\widehat{T}_{JK}$         & 962  & 88  & 96  & 1034 & 88  & 94  & 1054 & 763 & 764 \\
$\widehat{T}_{+1}$         & 1342 & 116 & 361 & 1245 & 59  & 252 & 1118 & 30  & 122 \\
$\widehat{T}_1$            & 796  & 64  & 213 & 835  & 35  & 168 & 884  & 19  & 117 \\
$\widehat{T}_2$            & 787  & 59  & 220 & 816  & 30  & 186 & 858  & 15  & 142 \\
\hline
$\widehat{T}_{\lambda}$    & 915  & 189 & 207 & 891  & 65  & 127 & 912  & 26  & 92 \\

\hline
\end{tabular}
\bigskip

\caption{Less uniform population: $p_i$'s $\sim U[0,1]$, $\gamma^2 \approx 0.3317$.}
\end{table}

\bigskip
\bigskip

\begin{table}[ht]
\centering

\begin{tabular}{|c || c|c|c || c|c|c || c|c|c ||}
\hline   
$T=1000$ & \multicolumn{3}{|c||}{$n=500$} & \multicolumn{3}{|c||}{$n=1000$} & \multicolumn{3}{|c||}{$n=2000$} \\
\cline{2-10}
                           & mean & std & MSE & mean & std & MSE & mean & std & MSE \\
\hline
\hline
$\widehat{T}_{TG}$         & 620  & 43  & 382 & 659  & 24  & 341 & 759  & 16  & 241 \\
$\widehat{T}_{CL}$         & 690  & 65  & 316 & 784  & 46  & 220 & 888  & 30  & 115 \\
$\widehat{T}_{JK}$         & 870  & 119 & 176 & 955  & 432 & 435 & 1027 & 661 & 662 \\
$\widehat{T}_{+1}$         & 1036 & 85  & 92  & 990  & 47  & 48  & 1013 & 30  & 30 \\ 
$\widehat{T}_1$            & 658  & 52  & 345 & 733  & 33  & 268 & 845  & 23  & 156\\
$\widehat{T}_2$            & 620  & 43  & 382 & 659  & 24  & 341 & 759  & 16  & 241 \\ 
\hline
$\widehat{T}_{\lambda}$    & 910  & 164 & 187 & 973  & 66  & 71  & 1001 & 42  & 42 \\

\hline
\end{tabular}
\bigskip

\caption{Non-uniform population: $p_i$'s $\sim Exp(1)$, $\gamma^2 \approx 0.9992$.}
\end{table}

\bigskip\bigskip

\begin{table}[ht]
\centering

\begin{tabular}{|c || c|c|c || c|c|c || c|c|c ||}
\hline   
$T=1000$ & \multicolumn{3}{|c||}{$n=500$} & \multicolumn{3}{|c||}{$n=1000$} & \multicolumn{3}{|c||}{$n=2000$} \\
\cline{2-10}
                           & mean & std & MSE & mean & std & MSE & mean & std & MSE \\
\hline
\hline
$\widehat{T}_{TG}$         & 192  & 10  & 808 & 228  & 8   & 772 & 261  & 7   & 738 \\
$\widehat{T}_{CL}$         & 262  & 26  & 737 & 346  & 28  & 654 & 416  & 29  & 583 \\
$\widehat{T}_{JK}$         & 326  & 486 & 830 & //   & //  & //  & //   & //  & //  \\
$\widehat{T}_{+1}$         & 271  & 19  & 729 & 304  & 15  & 696 & 334  & 14  & 666 \\
$\widehat{T}_1$            & 231  & 16  & 768 & 291  & 14  & 708 & 344  & 14  & 656 \\
$\widehat{T}_2$            & 192  & 10  & 808 & 228  & 8   & 772 & 261  & 7   & 738 \\   
\hline
$\widehat{T}_{\lambda}$    & 271  & 19  & 729 & 304  & 15  & 696 & 334  & 14  & 666 \\    

\hline
\end{tabular}
\bigskip

\caption{Extremely skewed population: $p_i$'s $\sim \Gamma(1,1)$, $\gamma^2 \approx 9.1289$.}
\end{table}
\bigskip
\bigskip

\begin{table}[h]\label{Tab_AlfE}
\centering
\begin{tabular}{|c || c|c|c || c|c|c || c|c|c ||}
\hline   
$T=26$ & \multicolumn{3}{|c||}{$n=15$} & \multicolumn{3}{|c||}{$n=25$} & \multicolumn{3}{|c||}{$n=50$} \\
\cline{2-10}
                           & media &  std & MSE & media & std & MSE & media & std & MSE \\
\hline
\hline
$\widehat{T}_{TG}$         &  19.8 & 10.3 &  70 & 18.6 & 3.6 & 56 &  19.5 & 2.5 & 43 \\
$\widehat{T}_{CL}$         &  22.7 & 13.3 &  71 & 20.7 & 5.6 & 56 &  21.6 & 4.0 & 37 \\
$\widehat{T}_{JK}$         &  19.4 &  6.9 &  60 & 23.0 & 9.8 & 58 &   // &  // & // \\
$\widehat{T}_{+1}$         &  33.3 & 20.3 &  109& 27.0 & 7.0 & 56 & 25.6 & 4.8 & 29 \\
\hline
$\widehat{T}_{\lambda}$    &  26.9 & 15.8 &  87 & 23.2 & 7.5 & 60 & 23.4 & 5.3 & 36 \\
\hline
\end{tabular}
\bigskip
\caption{Estimates for the 26 letters English alphabet  from
samples drawn from
  \cite{vocabolario1}; $\gamma^2 \approx 0.7029$ (see 
  \cite{LETTER1})  }
\end{table}

\begin{table}[h]
\centering
\begin{tabular}{|c || c|c|c || c|c|c || c|c|c ||}
\hline   
$T=21$ & \multicolumn{3}{|c||}{$n=15$} & \multicolumn{3}{|c||}{$n=25$} & \multicolumn{3}{|c||}{$n=50$} \\
\cline{2-10}
                           & media &  std & MSE & media & std & MSE & media & std & MSE \\
\hline
\hline
$\widehat{T}_{TG}$         &  16.0 &  6.9 & 62  & 15.7 &  3.4 & 43 & 16.7 &   2.0 & 31 \\
$\widehat{T}_{CL}$         &  18.4 &  9.8 & 71  & 17.7 &  5.4 & 45 & 18.5 &   3.3 & 27 \\
$\widehat{T}_{JK}$         &  16.9 &  6.3 & 51  & 19.6 &  8.2 & 65 &   // &    // & //  \\
$\widehat{T}_{+1}$         &  26.3 & 14.3 & 113 & 23.1 &  6.8 & 48 & 21.5 &   4.0 & 27 \\
\hline
$\widehat{T}_{\lambda}$    &  21.5 & 12.2 &  90 & 19.9 &  7.3 & 49 & 19.8 &   4.3 & 29 \\
\hline
\end{tabular}
\bigskip
\caption{Estimates for the 21 letters Italian alphabet  from
samples drawn from
 \cite{ADAMO}; $\gamma^2 \approx 0.5932$ (see 
  \cite{LETTER2}) }
\end{table}

\begin{table}[ht]
\centering

\begin{tabular}{|c|c || c|c|c  ||}
\hline   
Confidence level & Population $->$ &1&2&3 \\
\hline
\hline
$90\%$ \quad & fraction of hits      & 93\%  &92\%  &80\%  \\
 & average interval size       &1115  & 821  & 707  \\
\hline
\hline
$95\%$ \quad &fraction of hits         & 95\%  &98\%  &89\%    \\
 & average interval size           & 1225 & 889 & 827 \\
\hline
\hline
$99\%$ \quad &fraction of hits           & 97\%  &100\%  &98\%  \\
 & average interval size                &1520  & 1064  & 977 \\

\hline
\end{tabular}
\bigskip

\caption{Summary of confidence interval performances at the given confidence level from $\widehat T_{\widehat \lambda}$ by resampling.}
\end{table}

\vskip 5 in

\pagebreak

\newpage

\section{APPENDIX 1: The Bayesian approach}

By definition of the gamma  and beta functions
$
\Gamma(x)= \int_{0}^{+ \infty} e^{-t} t^{x-1} dt  \quad x>0
$ and  
\begin{displaymath}
\beta(x,y)=\int_{0}^{1} t^{a-1} (1-t)^{b-1} dt= \frac{\Gamma(x) \Gamma(y)}{\Gamma(x+y)},
\end{displaymath}
 taking  $z=y/(1-x)$ we get
\begin{displaymath}
\int_{0}^{1-x} y^a (1-x-y)^b dy=\int_{0}^{1} (1-x)^{a+b+1} z^a (1-z)^b dz =(1-x)^{a+b+1} \frac{\Gamma(a+1) \Gamma(b+1)}{\Gamma(a+b+2)}.
\end{displaymath}
Next, let $\rho_{n,T,\lambda}$ be the Bayesian posterior, given a 
sample with species records $m_1, \dots, m_N$, from a Dirichlet
prior with parameter $\lambda$ on
\begin{displaymath}
Q_T=\{ p=(p_1 \ldots p_{T-1}): p_k>0, \sum_{k=1}^{T-1} p_k \leq 1 \}.
\end{displaymath}
Note that $\rho_{n,T,\lambda}$ is invariant under permutation of the $p_k$'s, so it
is valid to express any result via a permutation of indices from a 
proven statement. Therefore, in the following Theorems it is
 sufficient to prove the results  for
some index $i$.

\begin{teo}\label{Ep}
For evey $\lambda \geq 1$ and for every  $i=1 \ldots T$, 
\begin{equation}
\EE_{\rho_{n,T,\lambda}}(p_k)=\frac{m_k+\lambda}{T\lambda+n}.
\end{equation}
\end{teo} 
\begin{proof} 
 For $i \in \{1 \ldots T-1 \}$ we have: 
\begin{eqnarray}
\EE_{\rho_{n,T,\lambda}}(p_i) & = & \int_{Q_T}p_i \rho_{n,T,\lambda}(d\mu) \nonumber \\
 & = & \frac{\int_{Q_T} p_1^{m_1+\lambda-1} \ldots p_i^{m_i+\lambda} \ldots (1-p_1- \ldots -p_{T-1})^{m_T+\lambda-1} dp_1 \ldots dp_{T-1}}{\int_{Q_T} p_1^{m_1+\lambda-1} \ldots p_i^{m_i+\lambda-1} \ldots (1-p_1- \ldots -p_{T-1})^{m_T+\lambda-1} dp_1 \ldots dp_{T-1}}. \nonumber
\end{eqnarray}
For $k=1 \ldots T$, let
\begin{eqnarray}
s_k&=&m_k+\lambda-1 \nonumber \\
 \widehat{s}_k&=&s_k+\delta(k,i)\nonumber
\end{eqnarray}
where $\delta$ is the Kronecker delta and for $k=1 \ldots T-1$ let
\begin{displaymath}
I(k)=\int_{Q_k} p_1^{s_1} \ldots p_{k}^{s_{k}} (1-p_1- \ldots -p_{k})^{s_T+ \ldots +s_{k+1}+T-k-1} dp_1 \ldots dp_{k}
\end{displaymath}
and
\begin{displaymath}
G(k)=\frac{\Gamma(s_k+1) \Gamma(s_T+ \ldots +s_{k+1}+T-k)}{\Gamma(s_T+ \ldots +s_{k+1}+s_k+T-k+1)}
\end{displaymath}
and let $\widehat{I}(k)$ and $\widehat{G}(k)$ be as the quantities without
hat but with $\widehat{s}_k$ replacing $s_k$, so that
 $$\EE_{\rho_{n,T,\lambda}}(p_i)=\frac{\widehat{I}(T-1)}{I(T-1)}.$$
Now we have 
\begin{eqnarray}
I(T-1) & = & \frac{\Gamma(s_{T-1}+1) \Gamma(s_T+1)}{\Gamma(s_T+s_{T-1}+2)} I(T-2)                          \nonumber \\
       & = & G(T-1)\frac{\Gamma(s_{T-2}+1) \Gamma(s_T+s_{T-1}+2)}{\Gamma(s_T+s_{T-1}+s_{T-2}+3)} I(T-3)    \nonumber \\
       & = & \prod_{k=1}^{T-1}G(k) = \frac{\Gamma(s_T+1) \ldots \Gamma(s_1+1)}{\Gamma(s_T+ \ldots +s_1+T)} \nonumber 
\end{eqnarray}
 and      
\begin{displaymath}
\widehat{I}(T-1)=\frac{\Gamma(\widehat{s}_T+1) \ldots \Gamma(\widehat{s}_1+1)}{\Gamma(\widehat{s}_T+ \ldots +\widehat{s}_1+T)} = \frac{\Gamma(s_T+1) \ldots \Gamma(s_i+2) \ldots \Gamma(s_1+1)}{\Gamma(s_T+ \ldots +s_1+T+1)}
\end{displaymath}
Therefore,
\begin{eqnarray}
\EE_{\rho_{n,T,\lambda}}(p_i) & = & \frac{\Gamma(s_i+2) \Gamma(s_T+ \ldots +s_1+T)}{\Gamma(s_i+1) \Gamma(s_T+ \ldots +s_1+T+1)} = \frac{s_i+1}{s_T+ \ldots s_1+T}\nonumber \\
															& = & \frac{m_i+\lambda}{m_1+ \ldots +m_T +T\lambda} = \frac{m_i+\lambda}{T\lambda+n}. \nonumber
\end{eqnarray}
It is easily verified that $\sum_{k=1}^{T}\frac{m_k+\lambda}{T\lambda+n}=1$.

Moreover, adding these values over the $T-N$ unobserved species we get an estimate
of $U$:
\begin{displaymath}
\widehat{U}_{+\lambda} = \EE_{\rho_{n,T,\lambda}}(U) = \EE_{\rho_{n,T,\lambda}}\Big( \sum_{m_i=0} p_i \Big) = \sum_{i=N+1}^{T}\EE_{\rho_{n,T,\lambda}}(p_i) = \frac{(T-N)\lambda}{T\lambda+n} 
\end{displaymath}
\end{proof}

\medskip
\begin{lem}\label{Epp}
For every $\lambda \geq 1$ and $i,j=1 \ldots T$ such that $i \neq j$,
\begin{equation}
\EE_{\rho_{n,T,\lambda}}(p_i p_j)=\frac{(m_i+\lambda)(m_j+\lambda)}{(T\lambda+n+1)(T\lambda+n)}
\end{equation}
\end{lem}
\begin{proof}
Following the proof of Theorem \ref{Ep} let, for $k=1 \ldots T$,
\begin{eqnarray}
s_k           & = & m_k+\lambda-1    \nonumber \\
\widehat{s}_k & = & s_k+\delta(i,k)+\delta(j,k), \qquad i \neq j ,
\quad 1 \leq i,j \leq T-1\nonumber 
\end{eqnarray}
Thus 
\begin{displaymath}
\EE_{\rho_{n,T,\lambda}}(p_i p_j)=\int_{Q_T} p_1 p_j  \rho_{n,T,\lambda}(d\mu)=\frac{\widehat{I}(T-1)}{I(T-1)}
\end{displaymath}
where
\begin{eqnarray}
I(T-1) & = & \frac{\Gamma(s_T+1) \ldots \Gamma(s_1+1)}{\Gamma(s_T+ \ldots +s_1+T)} \nonumber \\
\widehat{I}(T-1) & = & \frac{\Gamma(s_T+1) \ldots \Gamma(s_i+2) \ldots \Gamma(s_j+2) \ldots \Gamma(s_1+1)}{\Gamma(s_T+ \ldots +s_1+T+2)} \nonumber
\end{eqnarray}
Therefore
\begin{eqnarray}
\EE_{\rho_{n,T,\lambda}}(p_i p_j) & = & \frac{\Gamma(s_i+2)\Gamma(s_j+2)\Gamma(s_T+ \ldots +s_1+T)}{\Gamma(s_i+1)\Gamma(s_j+1)\Gamma(s_T+ \ldots +s_1+T+2)} \nonumber \\
& = & \frac{(s_i+1)(s_j+1)}{(\sum s_k+T)(\sum s_k+T+1)}\nonumber \\
& = & \frac{(m_i+\lambda)(m_j+\lambda)}{(T\lambda+n)(T\lambda+n+1)} \nonumber
\end{eqnarray}
\end{proof}

 \medskip

\begin{lem}\label{Ep2}
For every $\lambda \geq 1$ and for every $k=1 \ldots T,$
\begin{equation}
\EE_{\rho_{n,T,\lambda}}(p_k^2)=\frac{(m_k+\lambda+1)(m_k+\lambda)}{(T\lambda+n+1)(T\lambda+n)} 
\end{equation}
\end{lem}
\begin{proof}
As in Theorem \ref{Ep}, for $k=1 \ldots T$ and $i \in \{1 \ldots T-1\}$
let
\begin{eqnarray}
s_k           & = & m_k+\lambda-1    \nonumber \\
\widehat{s}_k & = & s_k+2\delta(k,i)  \nonumber 
\end{eqnarray}
So, 
$
\EE_{\rho_{n,T,\lambda}}(p_i^2)=\int_{Q_T} p_i^2 \quad \rho_{n,T,\lambda}(d\mu)=\frac{\widehat{I}(T-1)}{I(T-1)}
$
where
\begin{eqnarray}
I(T-1) & = & \frac{\Gamma(s_T+1) \ldots \Gamma(s_1+1)}{\Gamma(s_T+ \ldots +s_1+T)} \nonumber \\
\widehat{I}(T-1) & = & \frac{\Gamma(s_T+1) \ldots \Gamma(s_i+3) \ldots \Gamma(s_1+1)}{\Gamma(s_T+ \ldots +s_1+T+2)} \nonumber
\end{eqnarray}
Therefore, for $i=1, \dots, T-1$,
\begin{eqnarray}
\EE_{\rho_{n,T,\lambda}}(p_i^2) & = & \frac{\Gamma(s_i+3) \Gamma(s_T+ \ldots +s_1+T)}{\Gamma(s_i+1) \Gamma(s_T+ \ldots +s_1+T+2)} = \frac{(s_i+1)(s_i+2)}
{(\sum_{k=1}^T s_k+T)(\sum_{k=1}^T s_k+T+1)}\nonumber \\
															& = & \frac{(m_i+\lambda)(m_i+\lambda+1)}{(T\lambda+n)(T\lambda+n+1)} \nonumber
\end{eqnarray}
\end{proof}

\begin{lem}
If $q=\sum_{j \geq 0} j^2n_j = \sum_{k=1}^T m_k^2$ we have
\begin{equation}\label{P2}
\sum_{k=1}^{T} \EE_{\rho_{n,T,\lambda}}(p_k^2)=\frac{q+n(2\lambda+1)+T(\lambda^2+\lambda)}{(T\lambda+n+1)(T\lambda+n)}
\end{equation}
\end{lem}
\begin{proof}
We have
\begin{eqnarray}
\sum_{k=1}^{T} \EE_{\rho_{n,T,\lambda}}(p_k^2) & = & \sum_{k=1}^{T} \frac{(m_k+\lambda)(m_k+\lambda+1)}{(T\lambda+n)(T\lambda+n+1)} \nonumber \\
& = & \frac{\sum m_k^2+n(2\lambda+1)+T(\lambda^2+\lambda)}{(T\lambda+n+1)(T\lambda+n)} \nonumber \\
& = & \frac{q+n(2\lambda+1)+T(\lambda^2+\lambda)}{(T\lambda+n+1)(T\lambda+n)} \nonumber
\end{eqnarray}
\end{proof}

\section{APPENDIX 2: Some properties of the function defining $\lambda$}

Let $u= \widehat U$ and $v= \widehat V$. We consider now $u$ and $v$ as free
variables satisfying some requirements
satisfied by the values that, in fact, $\widehat U$ and $\widehat V$ take on in our estimation,
namely $\widehat U= \frac{n_1}{n}$ and $\widehat V= \widehat V_{GT}=\sum_{j \geq 1} \frac{j(j-1)n_j}{n(n-1)}$.

Also let $q =  vn(n-1)+n=\sum_{j>0}j^2n_j $. Then

\begin{lem}
For every sample, $\widehat{U} + \widehat{V} \leq 1$ 
\end{lem}
\begin{proof}
Since $q = \sum_{j>0}j^2n_j $ and 
$n = \sum_{j>0}j n_j$ we have that
\begin{eqnarray}
\widehat U + 
 \widehat V &=& \frac{n_1}{n} + \frac{q-n}{n(n-1)} \leq 1 \nonumber
 \end{eqnarray}
 is implied by
\begin{eqnarray}
  -n n_1 + n_1 -q + n^2 
&=& n_1 +(\sum_{j=1}^N j n_j)^2  -\sum_{j=1}^N n_j(jn_1+j^2 ) \nonumber\\
&=& \sum_{j=1}^N j^2 (n_j^2-n_j) +n_1 + \sum_{j, r=1, \dots,N, j \neq r} jn_j r n_r - n_1 \sum_{j=1}^N jn_j \nonumber\\
&\geq&   (n_1^2 - n_1)+ n_1+ n_1 \sum_{j=2}^N jn_j - n_1^2
\geq 0
\end{eqnarray}
\end{proof}

\begin{lem}
For every sample, $N +\widehat U -1-n \widehat U \geq 0$ 
\end{lem}
\begin{proof}
By definition of $\widehat U$ we have
$$
N +\widehat U -1-n \widehat U = N-n_1 +\frac{n_1}{n}-1,
$$
then either $n=n_1=N$ and the right hand side becomes $0$, or
$N- n_1\geq 1$ and the relation holds.
\end{proof}

\begin{lem}
For every sample, $(\widehat V nN-\widehat V N+N-n)n = qN-n^2 \geq 0$ 
\end{lem}
\begin{proof}
 Expressing $q,N$ and $n$ as function
of the $n_j$'s we get 
\begin{eqnarray}
q & = & \sum_{j>0}j^2n_j \nonumber \\
N & = & \sum_{j>0}n_j \nonumber \\
n & = & \sum_{j>0}j n_j.\nonumber
\end{eqnarray}
Then
\begin{eqnarray}
qN-n^2 & = & \Big(\sum_{j>0}j^2n_j\Big)\Big(\sum_{k>0}n_k\Big)-\Big(\sum_{j>0}jn_j\Big)^2 \nonumber \\
       & = & \Big(\sum_{j>0}j^2n_j^2+\sum_{j>0}\sum_{k \neq j}j^2n_jn_k\Big)-\Big(\sum_{j>0}j^2n_j^2+\sum_{j>0}\sum_{k \neq j}jn_j k n_k\Big) \nonumber \\
       & = & \sum_{j>0}\sum_{k \neq j}(j^2-jk)n_jn_k \nonumber \\
       & = & \sum_{j>0}\sum_{j<k}(j^2-jk+k^2-kj)n_jn_k \nonumber \\
       & = & \sum_{j>0}\sum_{j<k}(j-k)^2 n_jn_k \geq 0 \nonumber 
\end{eqnarray}
\end{proof}

Therefore, in the sequel we assume that $u$ and $v$ satisfy the following
relations:
\begin{eqnarray} 
0 &\leq& u \leq 1 \label{rel1} \\ 
0 &\leq& v \leq 1 \label{rel2} \\ 
 1 &\geq& u+v  \label{rel3} \\ 
 0 &\leq & N +u -1-n u \label{rel4} \\ 
 0 &\leq& v nN-v N+N-n \label{rel5}
\end{eqnarray}

\begin{teo}
Let
$$
f(\lambda)=v- \frac{vn(n-1)+n+n(2\lambda+1)+\lambda(\lambda+1)\frac{N+n u/\lambda}{1-u}}{[n+ \lambda \frac{N+n u/\lambda}{1-u}+1][n+ \lambda \frac{N+n u/\lambda}{1-u}]}
$$
we have
\begin{displaymath}
\widehat{\lambda}= \arg \inf_{\lambda \geq 1} \Big|f(\lambda)\Big| = \left\{ \begin{array}{ll}
 1 & \textrm{if $\beta_1 < \lambda_2$ and $1 \geq \lambda_2$, i.e. $\frac{2-v(N+1)}{2-v+vn} \leq u \leq 1-v$} \\
\lambda_2  & \textrm{if $\beta_1 < \lambda_2$ and $ \lambda_2 \geq 1$, i.e. $1-N v < u \leq \frac{2-v(N+1)}{2-v+vn}$} \\
\infty    & \textrm{if $\lambda_2 \leq \beta_1$, i.e. $0 \leq u \leq 1-N v$}.
\end{array} \right.
\end{displaymath}

where  $\beta_1= -\frac{n}{N}$ is the largest singularity of
 $f(\lambda)$ and
 $$
 \lambda_2= \frac{1-u-v+uv-uvn}{Nv+u-1} \label{LAMBDA33}.
 $$

\end{teo}

\begin{proof}
The equation $f(\lambda)=0$ has solutions:
\begin{eqnarray}
 \lambda_1 & = & \frac{-2n+ n u}{N} \leq 0\\
\lambda_2 & = & \frac{1-u-v+uv-uvn}{Nv+u-1} \label{LAMBDA3}
\end{eqnarray}
The root $\lambda_1$ is always non positive and thus it is not interesting and 
if 
\begin{equation} \label{concl1}
\lambda_2 = \frac{1-u-v+uv-uvn}{Nv+u-1} \geq 1, 
\end{equation}  
then $\lambda_2$ achieves
the required minimum.

To evaluate the other cases note that the function $f(\lambda)$ has two poles
\begin{eqnarray}
\beta_1 & = & -\frac{n}{N} 				\label{BETA1}				\\
\beta_2 & = & -\frac{n}{N}-\frac{1-u}{N}
\end{eqnarray}
and $\lambda_1 < \beta_2 < \beta_1$.
Moreover, 
$$\lim_{\lambda \rightarrow \beta_1^+} f(\lambda) =  \infty \cdot \textrm{sgn}(
\frac{(u-1)(v nN-v N+N-n)}{N^2}) = - \infty
$$
by \eqref{rel1} and \eqref{rel5}, and
\begin{equation} \label{concl2}
\lim_{\lambda \rightarrow +\infty} f(\lambda) =  \frac{ Nv+u-1}{N}.
\end{equation}

\medskip

We now verify that 
\begin{lem}
$f(\lambda)$ is increasing for $\lambda>\beta_1$. 
\end{lem}
\begin{proof}
Let $f'(\lambda)= \frac{(1-u) g(\lambda)}{(n+N \lambda)(1-u+n+N \lambda)}$.
Then 
\begin{equation} \label{app21}
\lim_{\lambda \rightarrow \beta_1^+} g(\lambda) = n(1-u)^2(v nN-v N+N-n)
> 0
\end{equation}
by \eqref{rel5}.
Note that $g$ satisfies
\begin{eqnarray}
g'(\lambda)&=& 2 N^2(N +u -1-n u) \lambda \nonumber \\
&&  + 2nN(-1-n+2N+u-Nu-Nv+n N v+N u v-n N u v)\nonumber
\end{eqnarray}
with the leading coefficient nonnegative by \eqref{rel4}. Therefore, if
$\lambda>\beta_1= -\frac{n}{N}$
\begin{eqnarray}
g'(\lambda)> 2nN(1-u)(v nN-v N+N-n) \geq 0 \nonumber
\end{eqnarray}
again by \eqref{rel5}.
Thus $g'>0$ for all $\lambda > \beta_1$ and, by \eqref{app21}, $g >0$ for 
all $\lambda > \beta_1$ and since the other factors in $f'$ are also 
positive, we have that $f' >0$ for all $\lambda > \beta_1$ as required.
\end{proof}

\medskip

Now there are three possibilities.
\begin{enumerate}

\item If $u \leq 1-Nv$ then from \eqref{concl2} and the above Lemma, it
follows that $f < 0 $ for all $\lambda > \beta_1$ and increasing, thus 
$$
\widehat \lambda=\arg \inf_{\lambda \geq 1} |f(\lambda)|=\arg \max{\lambda \geq 1} f=+ \infty.
$$

\item If $ 1-Nv < u <$ then from \eqref{concl1} $\lambda_2 \geq 1$
is equivalent to $u \leq \frac{2-v(N+1)}{2-v+vn}$, in which case
$
\widehat \lambda=\lambda_2.
$

\item If $ \frac{2-v(N+1)}{2-v+vn} < u$ then $\lambda_2 < 1$ and by the 
Lemma above 
$$
\widehat \lambda=\arg \inf_{\lambda \geq 1} |f(\lambda)|=\arg \min{\lambda \geq 1} f=1
$$

\end{enumerate}

The conditions on $u$ and $v$ are translated into those for $\lambda_2$ and
$\beta_1$ by direct calculation.

\end{proof}

\section*{Acknowledgements}

This work was done during visits by one of us (CCAS) to the
 Universit\`{a} di Roma, Tor
Vergata; Universit\`{a} di Milano-Bicocca;
and Universit\`{a} di Firenze. He takes pleasure in thanking those
universities for their warm hospitality and GNAMPA for 
its support.

\end{document}